\documentclass[a4]{article}
\pdfoutput=1
\usepackage{algorithm}
\usepackage{algpseudocode}
\usepackage{amsmath}
\usepackage{amsthm}
\usepackage{amsfonts} 
\usepackage{graphicx}
\usepackage{tikz}
\usepackage{bmpsize}

\usepackage{epstopdf}
\usetikzlibrary{arrows}

\newcommand{\Osymbol}{{O}}
\newcommand{\BO}[1]{\Osymbol\left(#1\right)}

\newcommand{\BOM}[1]{\Omega\left(#1\right)}

\newcommand{\BOx}[1]{\Osymbol(#1)}
\newcommand{\TOx}[1]{\tilde{\Osymbol}(#1)}

\newtheorem{theorem}{Theorem} 
\newtheorem{lemma}[theorem]{Lemma} 

\begin{document}
\title{Approximate Furthest Neighbor with Application to Annulus Query\footnote{The research leading to these results has received funding from the European Research Council under the European Union's Seventh Framework Programme(FP7/2007-2013) / ERC grant agreement no.~614331.}}

\author{Rasmus Pagh\footnote{pagh@itu.dk}, Francesco Silvestri\footnote{fras@itu.dk}, Johan Sivertsen\footnote{jovt@itu.dk} \&  Matthew Skala\footnote{mska@itu.dk}}

\maketitle
\begin{abstract}
Much recent work has been devoted to approximate nearest neighbor queries. 
Motivated by applications in recommender systems, we consider
\emph{approximate furthest neighbor} (AFN) queries and present a simple,
fast, and highly practical data structure for answering AFN queries in
high-dimensional Euclidean space.  The method builds on the technique of Indyk (SODA
2003), storing random projections to provide sublinear query time for
AFN\@. However, we introduce a different query algorithm, improving on Indyk's approximation 
factor and reducing the running time by a logarithmic factor.  We also present 
a variation based on a query-independent ordering of the database points; while 
this does not have the provable approximation factor of the query-dependent data 
structure, it offers significant improvement in time and space complexity.  
We give a theoretical analysis, and experimental results.
As an application, the query-dependent approach is used for deriving a data structure for the approximate annulus query problem, which is defined as follows: given an input set $S$ and two parameters $r>0$ and $w\geq 1$, construct a data structure that returns for each query point $q$ a point $p\in S$ such that the distance between $p$ and $q$ is at least $r/w$ and at most $w r$.\footnote{Formal publication DOI:  http://dx.doi.org/10.1016/j.is.2016.07.006}\footnote{This manuscript version is made available under the CC-BY-NC-ND 4.0 license http://creativecommons.org/licenses/by-nc-nd/4.0/}
\end{abstract}


\section{Introduction}

Similarity search is concerned with locating elements from a set $S$ that
are close to a given query $q$.  The query can be thought of as describing 
criteria we would like returned items to satisfy approximately.  For example, if a
customer has expressed interest in a product $q$, we may want to recommend
other, similar products.  However, we might not want to recommend products
that are \emph{too} similar, since that would not significantly increase the
probability of a sale.  Among the points that satisfy a near neighbor
condition (``similar''), we would like to return those that also satisfy a
furthest-point condition (``not too similar''), without explicitly computing
the set of all near neighbors and then searching it.  We refer to this
problem as the \emph{annulus query} problem.
We claim that an approximate solution to the annulus query problem can be
found by suitably combining Locality Sensitive Hashing (LSH), which is an
approximation technique commonly used for finding the nearest neighbor of a
query, with an approximation technique for furthest neighbor, which is the
main topic of this paper.  

The \emph{furthest neighbor} problem consists of
finding the point in an input set $S$ that maximizes the distance to a
query point $q$.  In this paper we investigate the
approximate furthest neighbor problem in $d$-dimensional Euclidean space
(i.e., $\ell^d_2$), with theoretical and experimental results.  We then
show how to cast one of our data structures to solve the annulus query problem.
As shown in the opening example, the furthest neighbor problem has been 
used in recommender systems to  create more diverse recommendations~\cite{said2013user,said2012increasing}. 
Moreover, the furthest neighbor is an important primitive in computational geometry, 
that has been used for computing the minimum spanning tree and the diameter of a
set of points~\cite{AgarwalMS92,Eppstein95}.

Our focus is on approximate solution because the exact version of the furthest neighbor problem would also
solve exact similarity search in $d$-dimensional Hamming space, and thus is
as difficult as that problem~\cite{Williams04,AhlePRS16}.  The reduction follows from the fact that the
complement of every sphere in Hamming space is also a sphere.  That limits
the hope we may have for an efficient solution to the exact version, so we
consider the \emph{$c$-approximate furthest neighbor} ($c$-AFN) problem
where the task is to return a point $x'$ with $d(q,x') \geq \max_{x\in S}
d(q,x) / c$, with $d(x,u)$ denoting the distance between two points.
We will pursue randomized solutions having a small
probability of not returning a $c$-AFN.  The success probability can be made
arbitrarily close to 1 by repetition.

We describe and analyze our data structures in Section~\ref{sec:alg}.  We
propose two approaches, both based on random projections but differing in 
what candidate points are considered at query time.
In the main query-dependent version the candidates will vary depending on the given query,
while in the query-independent version the candidates will be a fixed set.

The query-dependent data structure is presented in Section~\ref{sec:pq}.
It returns the
$c$-approximate furthest neighbor, for any $c>1$, with probability at least
$0.72$.  When the number of dimensions is $\BOx{\log n}$, our result
requires $\TOx{n^{1/c^2}}$ time per query and $\TOx{n^{2/c^2}}$ total space,
where $n$ denotes the input size.\footnote{The $\TOx{}$ notation omits polylog
terms.}  Theorem~\ref{thm:space} gives bounds in the general case.
This data structure is closely
similar to one proposed by Indyk~\cite{Indyk2003}, but we use a different approach for the  query algorithm. 

The query-independent data structure is presented in
Section~\ref{sub:query-ind}.  When the approximation factor is a constant
strictly between $1$ and $\sqrt{2}$, this approach requires $2^{\BO{d}}$
query time and space.  This approach is significantly faster than the
query dependent approach when the dimensionality is small.

The space requirements of our data structures are quite high: the query-independent data structure requires space exponential in the dimension, while 
the query-dependent one requires more than linear
space when $c<\sqrt{2}$. 
However, we claim that this bound cannot be significantly improved.
In Section~\ref{sec:lb} we show that any data structure that solves the $c$-AFN by storing a suitable subset of the input points
 must store at least $\min\{n, 2^{\BOM{d}}\}-1$
data points when $c<\sqrt{2}$.

Section~\ref{sec:exp} describes experiments on our data
structure, and some modified versions, on real and
randomly-generated data sets.  In practice, we can achieve approximation
factors significantly below the $\sqrt{2}$ theoretical result,
even with the query-independent version of the algorithm.
We can also
achieve good approximation in practice with significantly fewer projections
and points examined than the worst-case bounds suggested by the theory.  Our
techniques are much simpler to implement than existing methods for
$\sqrt{2}$-AFN, which generally require convex
programming~\cite{clarkson1995vegas,matouvsek1996subexponential}.  Our
techniques can also be extended to general metric spaces.

Having developed an improved AFN technique we return to the annulus
query problem in Section~\ref{sec:AAQ}.  We present a sublinear time
solution to the approximate annulus query problem based on combining our AFN
data structure with LSH techniques~\cite{Har-Peled2012}.

A preliminary version of our data structures for $c$-AFN appeared in the
proceedings of the 8th International Conference on Similarity Search and
Applications (SISAP)~\cite{PaghSSS15}.

\subsection{Related work}

\paragraph{Exact furthest neighbor}
In two dimensions the furthest neighbor problem can be solved in linear
space and logarithmic query time using point location in a furthest point
Voronoi diagram (see, for example,~de Berg et al.~\cite{CGbook08}).  However, the space usage of
Voronoi diagrams grows exponentially with the number of dimensions, making
this approach impractical in high dimensions.  More generally, an efficient data
structure for the \emph{exact} furthest neighbor problem in high dimension
would lead to surprising algorithms for satisfiability~\cite{Williams04}, so
barring a breakthrough in satisfiability algorithms we must assume that such
data structures are not feasible.
Further evidence of the difficulty of exact furthest neighbor is the following
reduction: Given a set
$S\subseteq \{-1,1\}^d$ and a query vector $q\in \{-1,1\}^d$, a
furthest neighbor (in Euclidean space) from $-q$ is a vector in $S$ of minimum
Hamming distance to $q$.  That is, exact furthest neighbor is at least as hard
as exact nearest neighbor in $d$-dimensional Hamming space, which is also 
believed to be hard for large $d$ and worst-case~\cite{Williams04}.

\paragraph{Approximate furthest neighbor}
Agarwal et al.~\cite{AgarwalMS92} proposes an algorithm for computing the $c$-AFN 
for \emph{all} points in a set $S$ in time $\BO{n/(c-1)^{(d-1)/2}}$ where $n=|S|$ and $1<c<2$.  
Bespamyatnikh~\cite{Bespam:Dynamic} gives a dynamic data structure for $c$-AFN.
This data structure relies on fair split trees and requires $\BO{1/(c-1)^{d-1}}$ time per query and $\BO{dn}$ space, with $1<c<2$.
The query times of both results exhibit an exponential dependency on the
dimension.  Indyk~\cite{Indyk2003} proposes the first approach avoiding this
exponential dependency, by means of multiple random projections of the data and query points to one
dimension.
More precisely, Indyk shows how to solve a
\emph{fixed radius} version of the problem where given a parameter $r$ the
task is to return a point at distance at least $r/c$ given that there exist
one or more points at distance at least $r$.  
Then, he gives a solution to the furthest
neighbor problem with approximation factor $c+\delta$, where $\delta > 0$ is a sufficiently small constant,
by reducing it to queries on many copies of that data structure.  The
overall result is
space $\tilde O(d n^{1+1/c^2})$ and query time $\tilde O(d n^{1/c^2})$, which improved the previous lower bound when $d=\BOM{\log n}$. 
The data structure presented in this paper shows that the same basic method, multiple random projections to one
dimension, can be used for solving $c$-AFN directly, avoiding the intermediate data structures for the fixed radius version.
Our result is then a simpler data structure that works for all radii and, being
interested in static queries, we are able to reduce the space to
$\TOx{dn^{2/c^2}}$.

\paragraph{Methods based on an enclosing ball}
Goel et al.~\cite{GIV01} show that a $\sqrt{2}$-approximate furthest
neighbor can always be found on the surface of the minimum enclosing ball of
$S$.  More specifically, there is a set $S^*$ of at most $d+1$ points from
$S$ whose minimum enclosing ball contains all of $S$, and returning the
furthest point in $S^*$ always gives a $\sqrt{2}$-approximation to the
furthest neighbor in $S$.  This method is \emph{query independent} in the
sense that it examines the same set of points for every query.  Conversely,
Goel et al.~\cite{GIV01} show that for a random data set consisting of $n$
(almost) orthonormal vectors, finding a $c$-approximate furthest neighbor
for a constant $c < \sqrt{2}$ gives the ability to find an
$O(1)$-approximate near neighbor.  Since it is not known how to do that in
time $n^{o(1)}$ it is reasonable to aim for query times of the form
$n^{f(c)}$ for approximation $c < \sqrt{2}$.

\paragraph{Applications in recommender systems}
Several papers on recommender systems have investigated the use
of furthest neighbor search~\cite{said2013user,said2012increasing}. 
The aim there was to use furthest neighbor search to create more diverse recommendations.
However, these papers do not address performance issues related to   
furthest neighbor search, which are the main focus of our paper. 
The data structures presented in this paper are intended to improve performance in recommender systems relying on furthest neighbor queries.
Other related works on recommender systems include those of
Abbar et al.~\cite{abbar2013real} and Indyk et
al.~\cite{indyk2014composable}, which use core-set techniques to return a
small set of recommendations no two of which are too close.  In turn,
core-set techniques also underpin works on approximating the minimum
enclosing ball~\cite{badoiu2008optimal,KMY03}.

\subsection{Notation}
\begin{figure}
\label{fig:annulus}
\caption{The $(r,w)$-annulus query.}
\centering
\vspace{2 mm}
\includegraphics[width=0.3\paperwidth,natwidth=800,natheight=800]{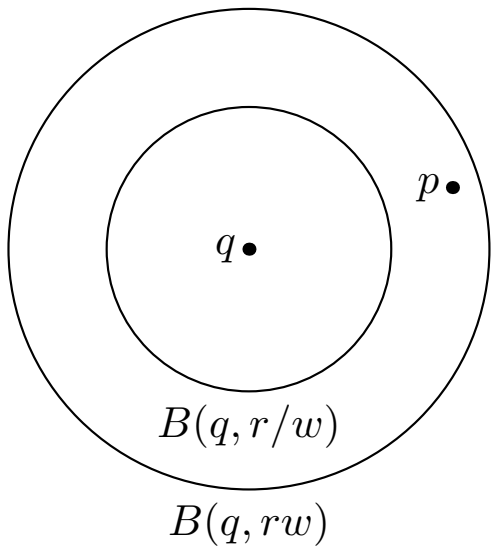}
\end{figure}

We use the following notation throughout:
\begin{itemize}
\item $B(x,r)$ for the set of all points in a ball of radius $r$ with center
  $x$.
\item $A(q,r,w)$ for the annulus between two balls, that is $A(q,r,w) = B(q,rw)-B(q,r/w)$. For an example, see Figure~\ref{fig:annulus}.
\item $[n]$ for the integers $1,..,n$.
\item $\arg\max_S^m f(x)$ for the set of $m$ elements from $S$ that have
  the largest values of $f(x)$, breaking ties arbitrarily.
 \item $N(\mu, \sigma^2)$ for the normal distribution with mean $\mu$ and variance $\sigma^2$.
\end{itemize}

\section{Algorithms and analysis}\label{sec:alg}

\subsection{Furthest neighbor with query-dependent candidates}\label{sec:pq}

Our data structure works by choosing a random line and storing the order of
the data points along it.  Two points far apart on the line are at least as
far apart in the original space.  So given a query we can find the points
furthest from the query on the projection line, and take those as candidates
to be the furthest point in the original space.  We build several such data
structures and query them in parallel, merging the results.

Given a set $S\subseteq \mathbb{R}^d$ of size $n$ (the input data),
let $\ell=2n^{1/c^2}$ (the number of random lines) and
$m =1+e^2 \ell \log^{c^2/2-1/3} n $ (the number of candidates to be examined
at query time), where $c>1$ is the desired
approximation factor.  We pick $\ell$ random vectors $a_1,\dots,a_\ell \in
\mathbb{R}^d$ with each entry of $a_i$ coming from the
standard normal distribution $N(0,1)$.  

For any $1\leq i \leq \ell$, we let $S_i = \arg\max^m_{x\in S} a_i\cdot x$
and store the elements of $S_i$ in sorted order according to the value
$a_i\cdot x$.  Our data structure for $c$-AFN consists of $\ell$ subsets
$S_1,\dots,S_\ell \subseteq S$, each of size $m$.  Since these subsets come
from independent random projections, they will not necessarily be disjoint
in general; but in high dimensions, they are unlikely to overlap very much. 
At query time, the algorithm searches for the furthest point from the query
$q$ among the $m$ points in $S_1,\dots,S_\ell$ that maximize $a_i
x- a_i q$, where $x$ is a point of $S_i$ and $a_i$ the random vector used
for constructing $S_i$.  The pseudocode is given in
Algorithm~\ref{alg:basic-query}.  We observe that although the data
structure is essentially that of Indyk~\cite{Indyk2003}, our technique
differs in the query procedure.

\begin{algorithm}
\caption{Query-dependent approximate furthest neighbor}\label{alg:basic-query}
\begin{algorithmic}[1]
\State initialize a priority queue of (point, integer) pairs, indexed by
  real keys
\For{$i=1$ to $\ell$}
  \State compute and store $a_i \cdot q$
  \State create an iterator into $S_i$, moving in decreasing
    order of $a_i\cdot x$
  \State get the first element $x$ from $S_i$ and advance the iterator
  \State insert $(x,i)$ in the priority queue with key
    $a_i\cdot x-a_i \cdot q$
\EndFor
\State $\mathit{rval} \leftarrow \bot$
\For{$j=1$ to $m$}
  \State extract highest-key element $(x,i)$ from the priority queue
  \If{$\mathit{rval} = \bot$ or $x$ is further than $\mathit{rval}$ from $q$}
    \State $\mathit{rval} \leftarrow x$
  \EndIf
  \State get the next element $x'$ from $S_i$ and advance the iterator
  \State insert $(x',i)$ in the priority queue with key
    $a_i\cdot x'-a_i \cdot q$
\EndFor
\State return $\mathit{rval}$
\end{algorithmic}
\end{algorithm}

Note that early termination is possible if $r$ is known at query time.

\paragraph{Correctness and analysis}
The algorithm examines distances to a set of at most $m$ points selected from the $S_i$, we will call the set $S_q$:
\[S_q\subseteq\cup_{i=1}^\ell S_i, ~ |S_q|\leq m.\]
We choose the name $S_q$ to emphasize that the set changes based on $q$.
Our algorithm succeeds if and only if $S_q$ contains a $c$-approximate
furthest neighbor. We now prove that this happens with constant
probability. 

We make use of the following standard lemmas that can be
found, for example, in the work of Datar et al.~\cite{Datar04} and Karger,
Motwani, and Sudan~\cite{KMS98}.

\begin{lemma}[See Section 3.2 of Datar et al.~\cite{Datar04}]\label{lem:n0}
For every choice of vectors $x,y \in \mathbb{R}^d$:
\begin{equation*}
\frac{a_i \cdot (x-y)}{\|x-y\|_2}\sim N(0,1).
\end{equation*}
\end{lemma}

\begin{lemma}[See Lemma 7.4 in Karger, Motwani, and Sudan~\cite{KMS98}]\label{lem:normalbound}
For every $t>0$, if $X\sim N(0,1)$ then
\begin{equation*}
\frac{1}{\sqrt{2\pi}}\cdot\left(\frac{1}{t}-\frac{1}{t^3}\right)\cdot e^{-t^2/2}\leq \Pr[X\geq t]\leq \frac{1}{\sqrt{2\pi}}\cdot\frac{1}{t}\cdot e^{-t^2/2}
\end{equation*}
\end{lemma}

The next lemma follows, as suggested by Indyk~\cite[Claims 2-3]{Indyk2003}.
\begin{lemma}\label{lem:prob}
Let $p$ be a furthest neighbor from the query $q$ with $r=\|p-q\|_2$, and
let $p'$ be a point such that $\|p'-q\|_2<r/c$.
Let $\Delta = rt/c$ with $t$ satisfying the equation
$e^{t^2/2}t^{c^2}=n/(2\pi)^{c^2/2}$
(that is, $t=\BO{\sqrt{\log n}}$).
Then, for a sufficiently large $n$, we have
\begin{gather*}
\Pr_a\left[a\cdot (p'-q)\geq \Delta\right]\leq \frac{\log^{c^2/2-1/3} n}{n}
\\
\Pr_a\left[a\cdot (p-q)\geq \Delta\right]\geq (1-o(1)) \frac{1}{n^{1/c^2}}\, .
\end{gather*}
\end{lemma}

\begin{proof}
Let $X\sim N(0,1)$. By Lemma~\ref{lem:n0} and the right part of
Lemma~\ref{lem:normalbound}, we have for a point $p'$ that
\begin{align*}
\Pr_a\left[a\cdot (p'-q)\geq \Delta\right]
&=\Pr_a\left[X\geq \Delta/\| p'-q\|_2\right]
  \leq \Pr_a\left[X\geq \Delta c/r\right] = 
  \Pr_a\left[X\geq t\right] 
  \\
&\leq \frac{1}{\sqrt{2\pi}} \frac{e^{-t^2/2}}{t}
  \leq \left(t \sqrt{2\pi}\right)^{c^2-1} \frac{1}{n} 
\leq \frac{\log^{c^2/2-1/3} n}{n}.
\end{align*}
The last step follows because $e^{t^2/2}t^{c^2}=n/(2\pi)^{c^2/2}$ implies that $t=\BO{\sqrt{\log n}}$, and holds for a sufficiently large $n$.
Similarly, by Lemma~\ref{lem:n0} and the left part of Lemma~\ref{lem:normalbound},
 we have for a furthest neighbor $p$ that
\begin{align*}
\Pr_a\left[a\cdot (p-q)\geq \Delta\right]
&=\Pr_a\left[X\geq \Delta/\| p-q\|_2\right]
  =\Pr_a\left[X\geq \Delta/r\right]
  =\Pr_a\left[X\geq t/c\right]\\
&\geq \frac{1}{\sqrt{2\pi}} \left(\frac{c}{t}-\left(\frac{c}{t}\right)^3\right){e^{-t^2/(2c^2)}}
  \geq (1-o(1))\frac{1}{n^{1/c^2}}.
\end{align*}
\qed
\end{proof}

\begin{theorem}
The data structure when queried by Algorithm~\ref{alg:basic-query}
returns a $c$-AFN of a given query with probability
$1-2/e^2>0.72$ in 
\begin{equation*}
\BO{n^{1/c^2}\log^{c^2/2-1/3}{n}(d + \log{n})}
\end{equation*}
time per query.  The data structure requires $\BOx{n^{1+1/c^2}(d + \log
n)}$ preprocessing time and total space
\begin{equation*}
  \BO{\min\left\{dn^{2/c^2}\log^{c^2/2-1/3}n, \,
    dn+n^{2/c^2}\log^{c^2/2-1/3}n\right\}} \, .
\end{equation*}
\end{theorem}

\begin{proof}
The space required by the data structure is the space required for storing
the $\ell$ sets $S_i$.  If for each set $S_i$ we store the $m\le n$ points and
the projection values, then $\BO{\ell m d}$ memory words are required.  On
the other hand, if pointers to the input points are stored, then the total
required space is $\BO{\ell m + nd}$.  The representations are equivalent,
and the best one depends on the value of $n$ and $d$.  The claim on space
requirement follows.  The preproceesing time is dominated by the
computation of the $n\ell$ projection values and by the sorting for
computing the sets $S_i$.  Finally, the query time is dominated by the at
most $2m$ insertion or deletion operations on the priority queue  
and the $md$ cost of searching   for the furthest neighbor, $\BO{m(\log{\ell}+d)}$.

We now upper bound the success probability. As in the statement of
Lemma~\ref{lem:prob}, we let $p$ denote a furthest neighbor from $q$,
$r=\|p-q\|_2$, $p'$ be a point such that $\|p'-q\|_2<r/c$, and $\Delta =
rt/c$ with $t$ such that $e^{t^2/2}t^{c^2}=n/(2\pi)^{c^2/2}$.  The query
succeeds if: (i) $a_i(p-q)\geq \Delta$ for at least one projection vector
$a_i$, and (ii) the (multi)set
$\hat{S}=\{p' | \exists i : a_i(p'-q)\geq \Delta,
\|p'-q\|_2<r/c\}$ contains at most $m -1$ points (i.e., there are at most $m
-1$ near points each with a distance from the query at least $\Delta$ in some
projection).  If (i) and (ii) hold, then the set of candidates examined by
the algorithm must contain the
furthest neighbor $p$ since there are at most $m-1$ points near to $q$ with
projection values larger than the maximum projection value of $p$.  Note
that we do not consider points at distance larger than $r/c$ but smaller
than $r$: they are $c$-approximate furthest neighbors of $q$ and can only
increase the success probability of our data structure.

By Lemma~\ref{lem:prob}, event (i) happens with probability $1/n^{1/c^2}$.
Since there are $\ell=2n^{1/c^2}$ independent projections, this event fails
to happen with probability at most $(1-1/n^{1/c^2})^{2n^{1/c^2}}\!\leq
1/e^2$.  For a point $p'$ at distance at most $r/c$ from $q$, the
probability that $a_i(p'-q)\geq \Delta$ is less than $(\log^{c^2/2-1/3}
n)/n$ for Lemma~\ref{lem:prob}.  Since there are $\ell$ projections of $n$
points, the expected number of such points is $\ell \log^{c^2/2-1/3} n$. 
Then, we have that $|\hat{S}|$ is greater than $m -1$ with probability at most
$1/e^2$ by the Markov inequality.  Note that a Chernoff bound cannot be used
since there exists a dependency among the projections onto the same random
vector $a_i$.  By a union bound, we can therefore conclude that the
algorithm succeeds with probability at least $1-2/e^2\geq 0.72$.
\qed \end{proof}

\subsection{Furthest neighbor with query-independent
candidates}\label{sub:query-ind}

Suppose instead of determining the candidates depending on the query point
by means of a priority queue, we choose a fixed candidate set to be
used for every query.  The $\sqrt{2}$-approximation
the minimum enclosing sphere is one example of such
a \emph{query-independent} algorithm.  In this section we consider a
query-independent variation of our projection-based algorithm.

During preprocessing, we choose $\ell$ unit vectors $y_1,y_2,\ldots,y_\ell$
independently and uniformly at random over the sphere
of unit vectors in $d$ dimensions.  We project the $n$ data points in $S$
onto each of these unit vectors and choose the extreme data point in each
projection; that is,
\begin{equation*}
  \left\{ \left. \arg \max_{x \in S} x\cdot y_i \right| i \in [\ell] \right\} \, .
\end{equation*}

The data structure stores the set of all data points so chosen;
there are at most $\ell$ of them, independent of $n$.
At query time, we check the query point $q$ against all the points we
stored, and return the furthest one.

To prove a bound on the approximation, we
will use the following result of B\"{o}r\"{o}czky and
Wintsche~\cite[Corollary~1.2]{Boroczky:Covering}.
Note that their notation differs from
ours in that they use $d$ for the dimensionality of the surface of the
sphere, hence one less than the dimensionality of the vectors, and $c$ for
the constant, conflicting with our $c$ for approximation factor.  We state the
result here in terms of our own variable names.

\begin{lemma}[See Corollary~1.2 in B\"{o}r\"{o}czky and
Wintsche~\cite{Boroczky:Covering}]\label{lem:cdc}
For any angle $\varphi$ with $0<\varphi<\arccos 1/\sqrt{d}$, in
$d$-dimensional Euclidean space, there exists a
set $V$ of at most $C_d(\varphi)$ unit vectors such that
for every unit vector $u$, there exists some $v \in V$ with the angle between
$u$ and $v$ at most $\varphi$, and 
\begin{equation}
  |V| \le
  C_d(\varphi) = \gamma \cos \varphi \cdot \frac{1}{\sin^{d+1} \varphi}
    \cdot {(d+1)}^{\frac{3}{2}} \ln (1+{(d+1)}\cos^2 \varphi) \, ,
  \label{eqn:cdc}
\end{equation}
where $\gamma$ is a universal constant.
\end{lemma}

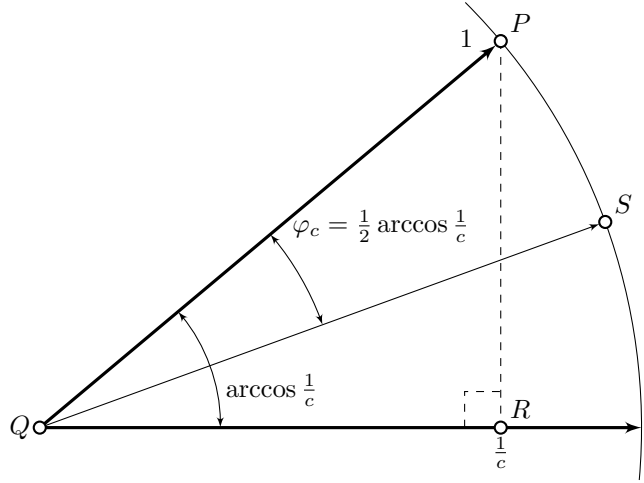
\begin{figure}
\centering
\begin{tikzpicture}[>=latex',scale=1.6]
  \draw[very thick,->] (0,0) -- (5,0);
  \draw[->] (0,0) -- (20:4.95);
  \draw[very thick,->] (0,0) -- (40:4.95);
  \draw[dashed] (40:5) -- (3.83,0);
  \draw[dashed] (3.53,0) -- (3.53,0.3) -- (3.83,0.3);
  \draw[<->] (1.5,0) arc[radius=1.5,start angle=0,end angle=40];
  \draw[<->] (20:2.5) arc[radius=2.5,start angle=20,end angle=40];
  \draw (-5:5) arc[radius=5,start angle=-5,end angle=45];
  \node[anchor=north] at (3.83,0) {$\frac{1}{c}$};
  \node[anchor=south east] at (40:4.8) {1};
  \node[anchor=west] at (12:1.5) {$\arccos \frac{1}{c}$};
  \node[anchor=south west] at (36:2.5)
    {$\varphi_c=\frac{1}{2}\arccos \frac{1}{c}$};
  \draw[black,thick,fill=white] (0,0) circle[radius=0.05];
  \draw[black,thick,fill=white] (20:5) circle[radius=0.05];
  \draw[black,thick,fill=white] (40:5) circle[radius=0.05];
  \draw[black,thick,fill=white] (3.83,0) circle[radius=0.05];
  \node[anchor=east] at (0,0) {$Q$};
  \node[anchor=south west] at (40:5) {$P$};
  \node[anchor=south west] at (20:5) {$S$};
  \node[anchor=south west] at (3.83,0) {$R$};
\end{tikzpicture}
\caption{Choosing $\varphi_c$.}\label{fig:varphi}
\end{figure}

Let $\varphi_c=\frac{1}{2}\arccos \frac{1}{c}$; that is half
the angle between two unit vectors whose dot product is $1/c$, as shown in
Figure~\ref{fig:varphi}.
Then by choosing $\ell={O(C_d(\varphi_c) \cdot \log C_d(\varphi_c))}$
unit vectors uniformly at
random, we will argue that with high probability we choose
a set of unit vectors such that
every unit vector has dot product at least $1/c$ with at least one of
them.  Then the data structure achieves $c$-approximation on all queries. 

\begin{theorem}
With $\ell=O(f(c)^d)$ for some function $f$ of $c$ 
and any $c$ such that $1<c<2$,
with high probability over the choice of the projection
vectors, the data structure returns a $d$-dimensional $c$-approximate
furthest neighbor on every query.
\end{theorem}

\begin{proof}
Let $\varphi_c=\frac{1}{2} \arccos \frac{1}{c}$.  Then, since $\frac{1}{c}$ is
between $\frac{1}{2}$ and $1$, we can apply the usual half-angle
formulas as follows:
\begin{gather*}
  \sin \varphi_c =
    \sin \frac{1}{2} \arccos \frac{1}{c}
  = \frac{\sqrt{1-\cos \arccos 1/c}}{\sqrt{2}}
  = \frac{\sqrt{1-1/c}}{\sqrt{2}} \\
  \cos \varphi_c =
    \cos \frac{1}{2} \arccos \frac{1}{c}
  = \frac{\sqrt{1+\cos \arccos 1/c}}{\sqrt{2}}
  = \frac{\sqrt{1+1/c}}{\sqrt{2}} \, .
\end{gather*}

Substituting into \eqref{eqn:cdc} from Lemma~\ref{lem:cdc}
gives
\begin{align*}
  C_d(\varphi_c)
  &= \gamma \frac{2^{d/2}\sqrt{1+1/c}}{(1-1/c)^{(d+1)/2}}
    {(d+1)}^{3/2} \ln \left(1+{(d+1)}\frac{1+1/c}{2}\right) \\
  &= O\left( \left( \frac{2}{1-1/c}\right) ^{(d+1)/2}
    d^{3/2} \log d \right) \, .
\end{align*}

Let $V$ be the set of $C_d(\varphi_c)$ unit vectors from
Lemma~\ref{lem:cdc}; every unit vector on the sphere is within angle at most
$\varphi_c$ from one of them.  The vectors in $V$ are the centres of a set
of spherical caps that cover the sphere.

Since the caps are all of equal size and they cover the sphere, there is
probability at least $1/C_d(\varphi_c)$ that a unit vector chosen uniformly
at random will be inside each cap.  Let $\ell= 2 C_d(\varphi_c) \ln
C_d(\varphi_c)$.  This $\ell = O(f(c)^d)$.  Then for each of the caps, the
probability none of the projection vectors $y_i$ is within that cap is
$(1-1/C_d(\varphi_c))^\ell$, which approaches $\exp (-2\ln C_d(\varphi_c)) =
(C_d(\varphi_c))^{-2}$.  By a union bound, the probability that every cap is
hit is at least $1-1/C_d(\varphi_c)$.  Suppose this occurs.

Then for any query, the vector between the query and the true furthest
neighbor will have angle at most $\varphi_c$ with some vector in $V$, and
that vector will have angle at most $\varphi_c$ with some projection vector
used in building the data structure.  Figure~\ref{fig:varphi} illustrates
these steps: if $Q$ is the query and $P$ is the true furthest neighbor, a
projection onto the unit vector in the direction from $Q$ to $P$ would give a
perfect approximation.  The sphere covering guarantees the existence of a
unit vector $S$ within an angle $\varphi_c$ of this perfect projection; and
then we have high probability of at least one of the random projections also
being within an angle $\varphi_c$ of $S$.  If that random projection returns
some candidate other than the true furthest neighbor, the worst case is if
it returns the point labelled $R$, which is still a $c$-approximation.
We have such approximations for all queries simultaneously with high
probability over the choice of the $\ell$ projection vectors. \qed
\end{proof}

Note that we could also achieve $c$-approximation deterministically, with
somewhat fewer projection vectors, by applying Lemma~\ref{lem:cdc} directly
with $\varphi_c=\arccos 1/c$ and using the centres of the covering caps
as the projection vectors instead of choosing them randomly.  That would
require implementing an explicit construction of the covering, however.
B\"{o}r\"{o}czky and
Wintsche~\cite{Boroczky:Covering} argue that their result is optimal to
within a factor $O(\log d)$, so not much asymptotic improvement is possible.

\subsection{A lower bound on the approximation factor}\label{sec:lb}

In this section, we show that a data structure aiming at an approximation
factor less than $\sqrt{2}$ must use space $\min\{n, 2^{\BOM{d}}\}-1$ on
worst-case data.
The lower bound holds for those data structures that compute the approximate
furthest neighbor by storing a suitable subset of the input points.

\begin{theorem}\label{thm:space}
Consider any data structure $\mathcal D$ that computes the $c$-AFN of an $n$-point input set $S\subseteq \mathbb{R}^d$ by storing a subest of the data set.
If $c=\sqrt{2}(1-\epsilon)$ with $\epsilon\in(0,1)$, then the algorithm must store at least  $\min\{n, 2^{\BOM{\epsilon^2 d}}\}-1$ points.
\end{theorem}
\begin{proof}
Suppose there exists a set $S'$ of size $r=2^{\BOM{\epsilon'^2 d}}$ such that for any $x\in S'$ we have $(1-\epsilon') \leq \|x\|_2^2\leq (1+\epsilon')$ and 
$x\cdot y\leq 2\epsilon'$, with $\epsilon'\in(0,1)$.
We will later prove that such a set exists.
We now prove  by contradiction that any data structure requiring less than $\min\{n, r\}-1$ input points cannot return a $\sqrt{2}(1-\epsilon)$-approximation.

Assume $n\leq r$. Consider the input set $S$ consisting of $n$ arbitrary points of $S'$ and set the query $q$ to $-x$, where $x$ is an input point not in the data structure.
The furthest neighbor is $x$ and it is at distance $\| x - (-x) \|_2\geq 2\sqrt{1-\epsilon'}$. 
On the other hand, for any point $y$ in the data structure, we get
\begin{equation*}
  \| y - (-x) \|_2 = \sqrt{\|x\|^2_2 + \|y\|^2_2 + 2 x\cdot y}
    \leq \sqrt{2(1+\epsilon') + 4\epsilon'}.
\end{equation*}
Therefore, the point returned by the data structure cannot be better than a $c'$ approximation with 
\begin{equation}\label{eq:approx}
c'= \frac{\| x - (-x) \|_2}{ \| y - (-x) \|_2} \geq 
\sqrt{2} \sqrt{\frac{1-\epsilon'}{1+3\epsilon'}}.
\end{equation}
The claim follows by setting $\epsilon'={(2\epsilon-\epsilon^2)/(1+3(1-\epsilon)^2)}$.

Assume now that $n> r$.  Without loss of generality, let $n$ be a multiple
of $r$.  Consider as input set the set $S$ containing $n/r$ copies of each
vector in $S'$, each copy expanded by a factor $i$ for any $i\in[n/r]$;
specifically, let $S=\{i x | x\in S', i\in[n/r] \}$.
By assumption, the data structure can store at most $r-1$ points and hence there exists a point $x\in S'$ such that $i x$ is not in the data structure for every $i\in[1, n/r]$.
Consider the query $q=-h x$ where $h=n/r$.
The furthest neighbor of $q$ in $S$ is $-q$ and it has distance $\| q - (-q) \|_2\geq 2h\sqrt{1-\epsilon'}$. 
On the other hand, for every point $y$ in the data structure, we get
\begin{equation*}
  \| y - (-hx) \|_2 = \sqrt{h^2\|x\|^2_2 + \|y\|^2_2 + 2 h x\cdot y}
  \leq \sqrt{2h^2(1+\epsilon') + 4h^2\epsilon'}.
\end{equation*}
We then get the same approximation factor $c'$ given in equation~\ref{eq:approx}, and the claim follows.
  
The existence of the set $S'$ of size $r$ follows from the
Johnson-Lindenstrauss lemma~\cite{Matousek:JL}. Specifically, consider an orthornormal base $x_1, \ldots x_r$ 
 of $\mathbb{R}^r$. 
Since $d=\BOM{\log r / \epsilon'^2}$, by the Johnson-Lindenstrauss
lemma there exists a linear map $f(\cdot )$ such that
$(1-\epsilon')\|x_i-x_j\|^2_2\leq \|f(x_i)-f(x_j)\|^2_2\leq
(1+\epsilon)\|x_i-x_j\|^2_2$ and $(1-\epsilon') \leq \|f(x_i)\|_2^2\leq
(1+\epsilon')$ for any $i,j$.  We also have that $f(x_i) \cdot
f(x_j)=(\|f(x_i)\|^2_2 + \|f(x_j)\|^2_2 - \|f(x_i)-f(x_j)\|^2_2)/2$, and
hence $-2\epsilon \leq f(x_i) \cdot f(x_j) \leq 2\epsilon$.  It then
suffices to set $S'$ to $\{f(x_1),\ldots, f(x_r)\}$.
\qed
\end{proof}

The lower
bound translates into the number of points that must be read by each query. 
However, this does not apply for query dependent data structures.


\section{Furthest neighbor experiments}\label{sec:exp}

We implemented several variations of furthest neighbor query in both the C
and F\# programming languages.  This code is available
online\footnote{https://github.com/johanvts/FN-Implementations}.  Our C
implementation is structured as an alternate index type for the SISAP C
library~\cite{SISAP:Library}, returning the furthest neighbor instead of the
nearest.

We selected five databases for experimentation:  the ``nasa''
and ``colors'' vector databases from the SISAP library; two randomly
generated databases of $10^5$ 10-dimensional vectors each, one using a
multidimensional normal distribution and one uniform on the unit cube; and
the MovieLens 20M dataset~\cite{Harper:MovieLens}.  The
10-dimensional random distributions were intended to represent
realistic data, but their intrinsic dimensionality as measured by the $\rho$
statistic of Ch\'{a}vez and Navarro~\cite{Chavez:Intrinsic} is significantly
higher than what we would expect to see in real-life applications.

\begin{figure}
\input{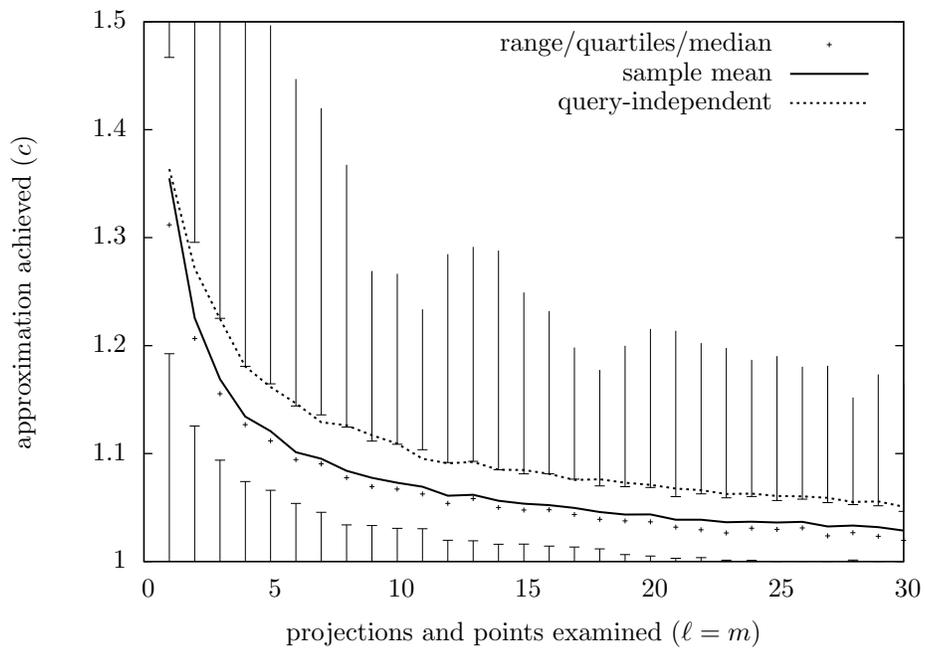}
\caption{Experimental results for 10-dimensional uniform distribution}
\label{fig:uniform}
\end{figure}

\begin{figure}
\input{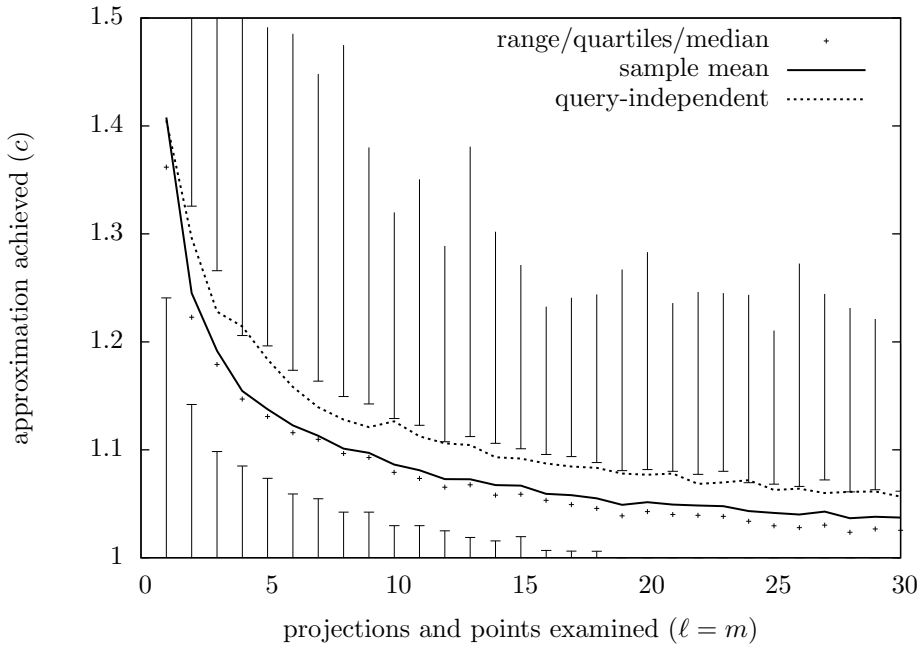}
\caption{Experimental results for 10-dimensional normal distribution}
\label{fig:normal}
\end{figure}

\begin{figure}
\input{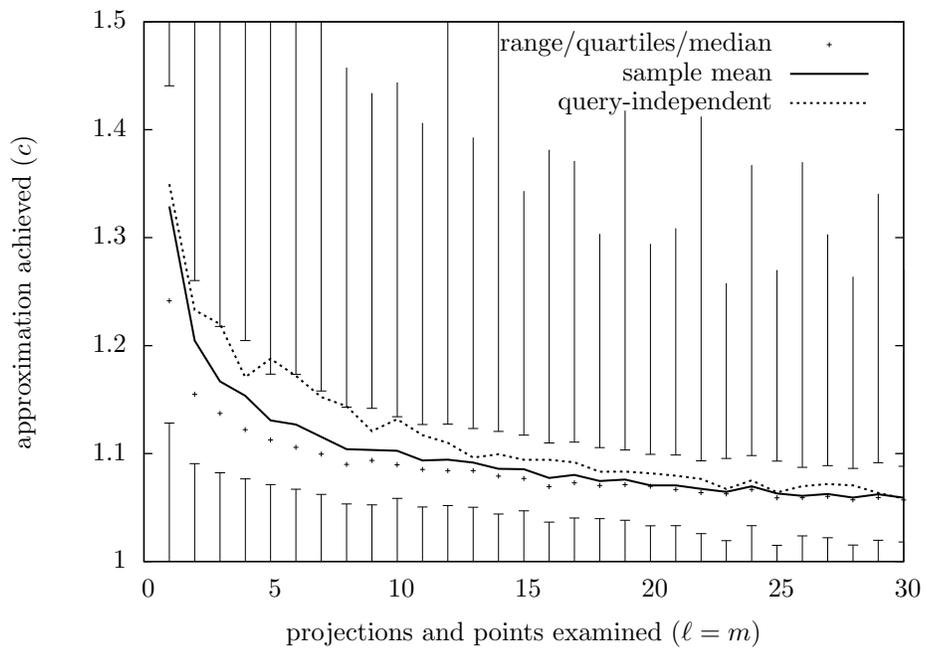}
\caption{Experimental results for SISAP nasa database}
\label{fig:nasa}
\end{figure}

\begin{figure}
\input{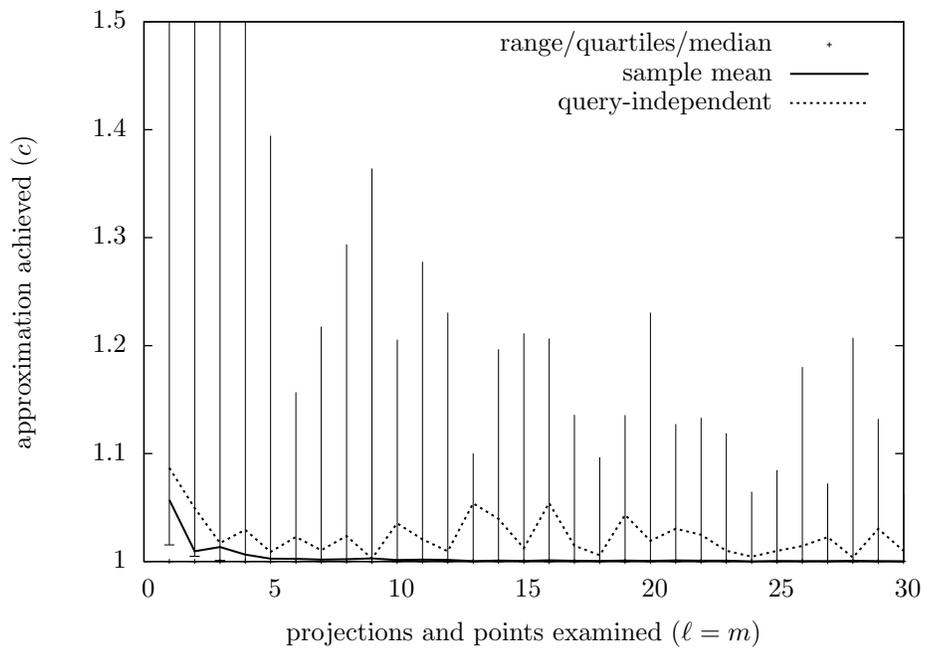}
\caption{Experimental results for SISAP colors database}
\label{fig:colors}
\end{figure}

\begin{figure}
\input{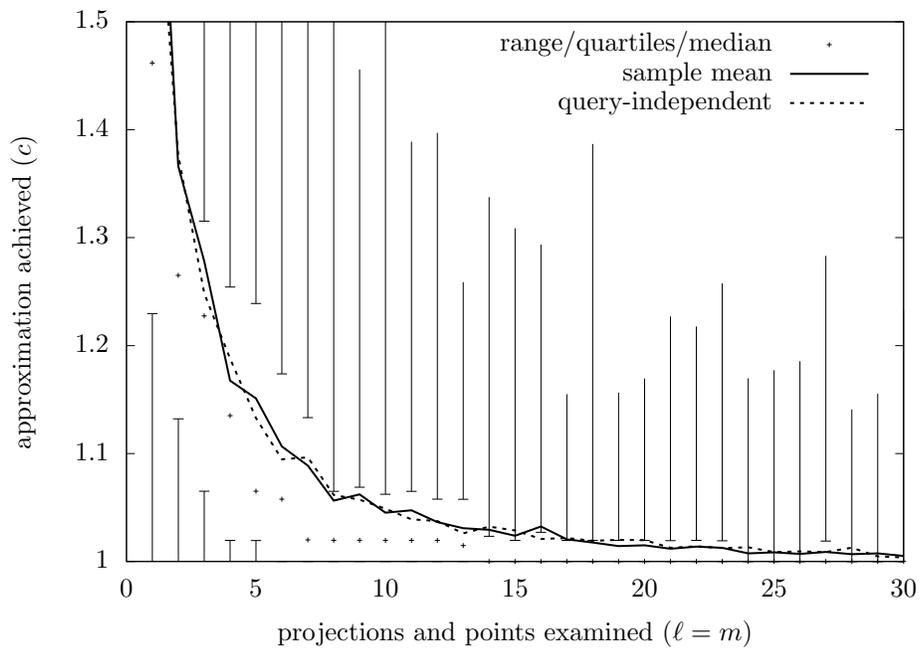}
\caption{Experimental results for MovieLens 20M database}
\label{fig:movies}
\end{figure}

For each database and each choice of $\ell$ from 1 to 30 and
$m$ from $1$ to $4\ell$, we made 1000 approximate furthest neighbor
queries.  To provide a representative sample over the randomization of both
the projection vectors and the queries, we used 100 different
seeds for generation of the projection vectors, and did 10 queries (each
uniformly selected from the database points) with each seed.  We computed
the approximation achieved, compared to the true furthest neighbor found by
brute force, for every query. The resulting distributions are summarized in
Figures~\ref{fig:uniform}--\ref{fig:movies}.

We also ran some experiments on higher-dimensional random vector
databases (with 30 and 100 dimensions, in particular) and saw approximation
factors very close to those achieved for 10 dimensions.

\paragraph{$\ell$ vs.\ $m$ tradeoff}

The two parameters $\ell$ and $m$ both improve the approximation as they
increase, and they each have a cost in the time and space bounds. 
The best tradeoff is not clear from the analysis. 
We chose $\ell=m$ as a typical value, but we
also collected data on many other parameter choices. 

\begin{figure}
\input{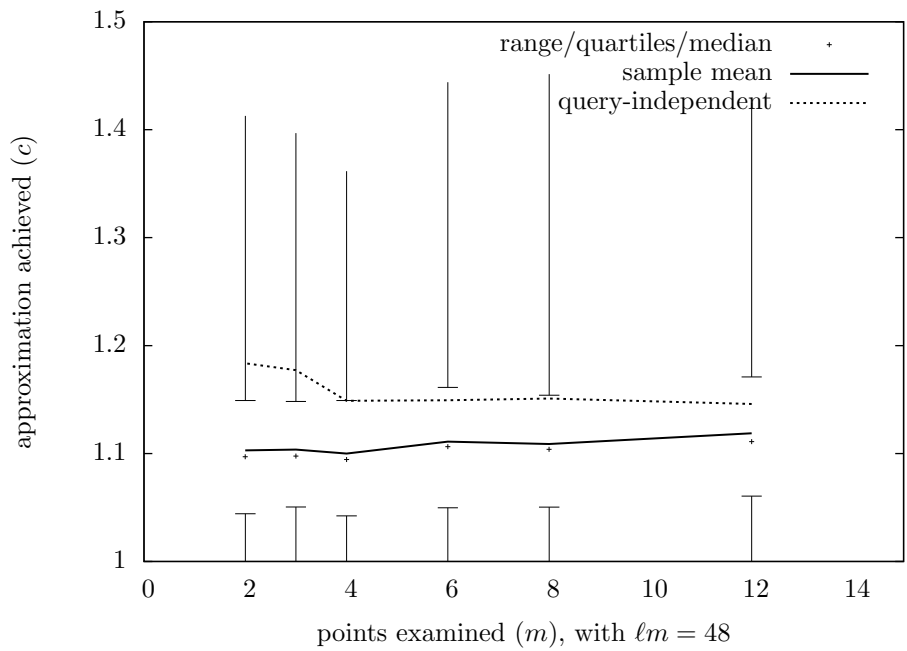}
\caption{The tradeoff between $\ell$ and $m$ on 10-dimensional normal
vectors}
\label{fig:tradeoff}
\end{figure}

Figure~\ref{fig:tradeoff} offers some insight into the tradeoff:
since the cost of doing a query is roughly proportional to both $\ell$ and
$m$, we chose a fixed value for their product, $\ell \cdot m=48$, and
plotted the approximation results in relation to $m$ given that, for the
database of normally distributed vectors in 10 dimensions.
As the figure shows, the approximation factor does not change
much with the tradeoff between $\ell$ and $m$.

\paragraph{Query-independent ordering}

The furthest-neighbor algorithm described in Section~\ref{sec:pq} examines
candidates for the furthest neighbor in a \emph{query dependent} order.
In order to compute the order for arbitrary queries, we must store
$m$ point IDs for each of the $\ell$ projections, and use a priority queue
data structure during query, incurring some costs in both time and space. 
It seems intuitively reasonable that the search will usually examine points
in a very similar order regardless of the query: first those that are
outliers, on or near the convex hull of the database, and then working its
way inward.  

We implemented a modified version of the algorithm in which the index stores
a single ordering of the points.  Given a set $S\subseteq \mathbb{R}^d$ of
size $n$, for each point $x \in S$ let
$\mathit{key}(x)=\max_{i \in 1\ldots\ell} a_i\cdot x$.  The key for each
point is its greatest projection value on any of the $\ell$
randomly-selected projections.  The data structure stores points (all of
them, or enough to accomodate the largest $m$ we plan to use) in order of
decreasing key value:  $x_1$, $x_2$, $\ldots$ where $\mathit{key}(x_1) \ge
\mathit{key}(x_2) \ge \cdots$.  Note that this is not the
same query-independent data structure discussed in
Section~\ref{sub:query-ind}; it differs both in the set of points
stored and the order of sorting them.

The query 
examines the first $m$ points in
the \emph{query independent} ordering and returns the one furthest from the
query point.  Sample mean approximation factor for this algorithm in our
experiments is shown by the dotted lines in
Figures~\ref{fig:uniform}--\ref{fig:tradeoff}.

\begin{algorithm}[t]
\caption{Query-independent approximate furthest
neighbor}\label{alg:independent}
\begin{algorithmic}[1]
\State $\mathit{rval} \leftarrow \bot$
\For{$j=1$ to $m$}
  \If{$\mathit{rval} = \bot$ or $x_j$ is further than $\mathit{rval}$ from $q$}
    \State $\mathit{rval} \leftarrow x_j$
  \EndIf
\EndFor
\State return $\mathit{rval}$
\end{algorithmic}
\end{algorithm}

\paragraph{Variations on the algorithm}

We have experimented with a number of practical improvements to the
algorithm.  The most significant is to use the rank-based \emph{depth} of
projections rather than the projection value.  In this variation we sort the
points by their projection value for each $a_i$.  The first and last point
then have depth 0, the second and second-to-last have depth 1, and so on up
to the middle at depth $n/2$.  We find the minimum depth of each point over
all projections and store the points in a query independent order using the
minimum depth as the key.  This approach seems to give better results in
practice.  A further improvement is to break ties in the minimum depth by
count of how many times that depth is achieved, giving more priority to
investigating points that repeatedly project to extreme values.  Although
such algorithms may be difficult to analyse in general, we give some results
in Section~\ref{sub:query-ind} for the case where the data structure
stores exactly the one most extreme point from each projection.

The number of points examined $m$ can be chosen per query and even
during a query, allowing for interactive search.  After returning the best
result for some $m$, the algorithm can continue to a larger $m$ for a
possibly better approximation factor on the same query.  The smooth tradeoff
we observed between $\ell$ and $m$ suggests that choosing an $\ell$ during
preprocessing will not much constrain the eventual choice of $m$.

\paragraph{Discussion}

The main experimental result is that the algorithm works very well for the
tested datasets in terms of returning good approximations of the furthest
neighbor.  Even for small $\ell$ and $m$ the algorithm returns good
approximations.  Another result is that the query independent variation of
the algorithm returns points only slighly worse than the query dependent. 
The query independent algorithm is simpler to implement, it can be queried
in time $\BO{m}$ as opposed to $\BO{m \log{\ell+m}}$ and uses only $\BO{m}$
storage.  In many cases these advances more than make up for the slightly
worse approximation observed in these experiments.  However, by
Theorem~\ref{thm:space}, to guarantee $\sqrt{2}-\epsilon$ approximation the
query-independent ordering version would need to store and read $m=n-1$
points.

In data sets of high intrinsic dimensionality, the furthest point from a
query may not be much further than any randomly selected point, and we can
ask whether our results are any better than a trivial random selection from
the database.  The intrinsic dimensionality statistic $\rho$ of Ch{\'a}vez
and Navarro~\cite{Chavez:Intrinsic} provides some insight into this
question.  Note that instrinsic dimensionality as measured by $\rho$ is not
the same thing as the number of coordinates in a vector. For real data sets it is often much smaller than that.  Intrinsic dimensionality also applies to
data sets that are not vectors and do not have coordinates.  Skala proves
a formula for the value of $\rho$ on a multidimensional normal
distribution~\cite[Theorem~2.10]{Skala:Dissertation}; it is
$9.768\ldots$ for the 10-dimensional distribution used in
Figure~\ref{fig:normal}.  With the definition $\mu^2/2\sigma^2$, this means
the standard deviation of a randomly selected distance will be about 32\% of
the mean distance.  Our experimental results come much closer than that to
the true furthest distance, and so are non-trivial.

The concentration of distances in data sets of high intrinsic dimensionality
reduces the usefulness of approximate furthest neighbor.  Thus, although we
observed similar values of $c$ in higher dimensions to our 10-dimensional
random vector results, random vectors of higher dimension may represent a
case where $c$-approximate furthest neighbor is not a particularly
interesting problem.  However, vectors in a space with many dimensions
but low intrinsic dimensionality, such as the colors database, are
representative of many real applications, and our algorithms performed well
on such data sets.

The experimental results on the MovieLens 20M data
set~\cite{Harper:MovieLens}, which were not included in the conference
version of the present work, show some interesting effects resulting from
the very high nominal (number of coordinates) dimensionality of this data
set.  The data set consists of 20000263 ``ratings,'' representing the
opinions of 138493 users on 27278 movies.  We treated this as a database of
27278 points (one for each movie) in a 138493-dimensional Euclidean space,
filling in zeroes for the large majority of coordinates where a given user
did not rate a given movie.  Because of their sparsity, vectors in this data
set usually tend to be orthogonal, with the distance between two simply
determined by their lengths.  Since the vectors' lengths vary over a wide
range (length proportional to number of users rating a movie, which varies
widely), the pairwise distances also have a large variance, implying a
low intrinsic dimensionality.  We measured it as $\rho=0.263$.

The curves plotted in Figure~\ref{fig:movies} show similar behaviour to that of 
the random distributions in Figures~\ref{fig:uniform}
and~\ref{fig:normal}.  Approximation factor improves rapidly with more
projections and points examined, in the same pattern, but to a greater
degree, as in the 10-coordinate vector databases, which have higher
intrinsic dimensionality.  However, here there is no noticeable penalty for
using the query-independent algorithm.  The data set appears to be dominated
(insofar as furthest neighbours are concerned) by a few extreme outliers:
movies rated very differently from any others.  For almost any query, it is
likely that one of these will be at least a good approximation of the true
furthest neighbour; so the algorithm that identifies a set of outliers in
advance and then chooses among them gives essentially the same results as
the more expensive query-dependant algorithm.


\section{Annulus query}
\label{sec:AAQ}

In this section we return to the problem of annulus query. Using the AFN data structure in combination with LSH techniques we present a sub-linear time data structure for solving the approximate annulus query problem (AAQ) with constant failure probability in Euclidean space.
Let's begin by defining the exact and approximate annulus query problem:

\emph{Annulus query}: Consider a set of points $S$ in $(X,D)$ and $r>0,w>1$.
The exact $(r,w)$-annulus query is defined as follows:
Given a query point $q$, return a point $p\in S \cap A(q,r,w)$.
That is, we search for $p\in S$ such that $r/w \le D(p,q) \le wr$.
If no such point exists in $S$ the query returns null. An alternative 
definition returns \emph{all} points in $S \cap A(q,r,w)$, but we will focus our attention on the definition above.

\emph{Approximate annulus query}: For a set of points $S$ in $(X,D)$, $r>0$ and $c,w>1$.
The $(c,r,w)$-approximate annulus query (AAQ) is defined as follows:
Given a query point $q$, if there exists $p\in S \cap A(q,r,w)$, then
return a point $\hat{p}\in S \cap A(q,r,cw)$.
If no such $p$ exists we can return either null or any point within $A(q,r,cw)$. 

 \subsection{Solving the $(c,w,r)$-AAQ}
We now show how to solve the $(c,w,r)$-AAQ with constant failure probability in $\mathbb{R}^d$ by combining the furthest neighbor technique with locality sensitive hashing methods \cite{Har-Peled2012}.
Consider an LSH function family $\mathcal{H}=\{\mathbb{R}^d\rightarrow U\}$. We say that $\mathcal{H}$ is $(r_1,r_2,p_1,p_2)$-sensitive for $(\mathbb{R}^d,\ell _2)$ if:
\begin{enumerate}
\item$\Pr_{\mathcal{H}}[h(q)=h(p)]\geq p_1 \text{ when } \|p-q\|_2\leq r_1$
\item$\Pr_{\mathcal{H}}[h(q)=h(p)]\leq p_2 \text{ when } \|p-q\|_2> r_2$
\end{enumerate}

\begin{theorem}
  \label{thm:aaq-structure}
  Consider a $(wr,wcr,p_1,p_2)$-sensitive hash family $\mathcal{H}$ for $(\mathbb{R}^d,l_2)$ and let $\rho = \frac{\log 1/p_1}{\log 1/p_2}$.
  For any set $S\in\mathbb{R}^d$ of at most $n$ points there exists a data structure for $(c,w,r)$-AAQ such that:
  \begin{itemize}
    \item Queries can be answered in time $\BO{dn^{1/c^2}\log^{(1-1/c^2)/2}{n}}$.
    \item The data structure takes space $\BO{n^{2(\rho+1/c^2)}\log^{1-1/c^2}{n}}$ in addition to storing $S$.
  \end{itemize}

The failure probability is constant and can be reduced to any $\delta>0$ by increasing the space and time cost by a constant factor.
\end{theorem}

We will now give a description of such a data structure and then prove that it has the properties stated in Theorem
\ref{thm:aaq-structure}.

\subsection{Annulus query data structure}
\label{AAQ:datastructure}
Let $k,\ell$ and $L$ be integer parameters to be chosen later
We construct a function family $\mathcal{G}={g: \mathbb{R}^d \rightarrow U^k}$ by concatenating $k$ members of $\mathcal{H}$. Choose $L$ functions $g_1,..,g_L$ from $\mathcal{G}$ and pick $\ell$ random vectors $a_1,..,a_\ell\in\mathbb{R}^d$ with entries sampled independently from $\mathcal{N}(0,1)$.

\subsubsection{Preprocessing}
\begin{figure}
\caption{Illustration of a bucket for $\{x_1,x_2,x_3,x_5\}\subset S$. $\ell=3$.}
\label{fig:bucket-contents}
\centering
\vspace{2 mm}
\includegraphics[width=0.3\paperwidth,natwidth=800,natheight=800]{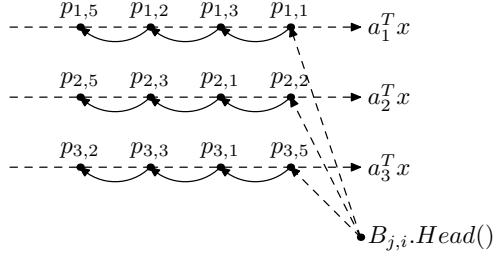}
\end{figure}

During  preprocessing, all points $x\in S$ are hashed with each of the functions $g_1,..,g_L$.
We say that a point $x$ is in a bucket $B_{j,i}$ if $g_j(x)=i$.
For every point $x\in S$ the $\ell$ dot product values $a_i\cdot x$ are calculated.
These values are stored in the bucket along with a reference to $x$.
Each bucket consists of $\ell$ linked lists, list $i$ containing the entries sorted on $a_i\cdot x$, decreasing from the head of the list.
See Figure~\ref{fig:bucket-contents} for an illustration where $p_{i,j}$ is the tuple $(a_i\cdot x_j,\text{ref}(x_j))$.
A bucket provides constant time access to the head of each list. Only non-empty buckets are stored.

\subsubsection{Querying}
For a given query point $q$ the query procedure can be viewed as building the set $S_q$ of points from $S$ within $B(q,rcw)$ with the largest $a_{i\in[\ell]}\cdot (p-q)$ values and computing the distances between $q$ and the points in $S_q$.
At query time $q$ is hashed using $g_1,..,g_L$.
From each bucket $B_{j,g_j(q)}$ the top pointer is selected from each list.
The selected points are then added to a priority queue with priority $a_i\cdot (p-q)$.
This is done in $\mathcal{O}(L\ell)$ time.
Now we begin a cycle of adding and removing elements from the priority queue. 
The largest priority element is dequeued and the predecessor link is followed and the returned pointer added to the queue.
If the pointer just visited was the last in its list, nothing is added to the queue. 
If the priority queue becomes empty the algorithm fails. 
Since $r$ is known at query time in the $(c,w,r)$-AAQ it is possible to terminate the query procedure as soon as some point within the annulus is found. 
Note that this differs from the general furthest neighbor problem. 
For the analysis however we will consider the worst case where only the last element in $S_q$ lies in the annulus and bound $|S_q|$ to achieve constant success probability. 

\begin{proof}
\label{prf.1}
  Fix a query point $q$.
  By the problem definition, we may assume $|S \cap A(q,r,w)|\geq1$.
  Define $S_q\subseteq S$ to be the set of candidate points for which the data structure described in section \ref{AAQ:datastructure} calculates the  distance to $q$ when queried. The correctness of the algorithm follows if $|S_q \cap A(q,r,cw)|\geq1$. 

  To simplify the notation let $P_{\text{near}} = S\cap B(q,r/(cw))$ and $P_{\text{far}}=S-B(q,r/w)$.
  Points in the these two sets have useful properties. 
  Let $t$ be the solution to the equality:
  \begin{equation*}
    \frac{1}{\sqrt{2\pi}}\frac{e^{\frac{-t^2}{2}}}{t}=\frac{1}{n}
  \end{equation*}

  If we set $\Delta=\frac{rt}{cw}$, we can use the ideas from Lemma \ref{lem:prob} to conclude that:

  \begin{equation*}
    \label{pb:near}
    \Pr[a_i(p-q)\geq\Delta]\leq\frac{1}{n}\text{, for }p\in P_{\text{near}}
  \end{equation*}
  Also, for $p\in P_{\text{far}}$ the lower bound gives:
  \begin{equation*}
    \Pr[a_i(p-q)\geq\Delta]\geq\frac{1}{(2\pi)^{(1-1/c2)/2}}n^{-1/c^2}t^{(1-1/c^2)}\left(1-\frac{c^2}{t^2}\right)
  \end{equation*}
  By definition, $t\in\BOx{\sqrt{\log{n}}}$, so for some function $\phi\in\BOx{n^{1/c^2}\log^{(1-1/c^2)/2}{n}}$ we get:
  \begin{equation*}
    \label{pb:far}
    \Pr[a_i(p-q)\geq\Delta]\geq\frac{1}{\phi}\text{, for }p\in P_{\text{far}}.
  \end{equation*}
  Now for large $n$,
  let $D$ be the set of points that hashed to the same bucket as $q$ for at least one hash function and projected above $\Delta$ on at least one projection vector.
  \begin{equation*}
    D=\{x\in S|\exists j,i:g_j(x)=g_j(q) \text{ and } a_i\cdot(x-q)\geq\Delta\}
  \end{equation*}  
  Let $\ell = 2\phi,m=1+e^2\ell$ and $L=\lceil n^\rho/p_1\rceil$.
  Using the probability bound (\ref{pb:near}) we see that $\text{E}[|D\cap P_{\text{near}}|]\leq\frac{1}{n}n\ell=\ell$.
  So $\Pr[|D\cap P_{\text{near}}|\geq m] < 1/e^2$ by Markov's inequality. 
  By a result of Har-Peled, Indyk, and Motwani~\cite[Theorem
  3.4]{Har-Peled2012}, the total number of points from $S-B(q,rcw)$ across all $g_i(q)$ buckets is at most $3L$ with probability at least $2/3$. So $\Pr[|D-B(q,rcw)>3L] < 1/3$.
  This bounds the number of too far and too near points expected in $D$.
  $$\Pr[|D\setminus A(q,r,cw)|\geq m +3L]\leq 1/3+e^{-2}$$
  By applying~\cite[Theorem
  3.4]{Har-Peled2012} again, we get that
  for each $x \in A(q,r,w)$ there exists $i\in[L]$ such that $g_i(x) = g_i(q)$ with probability at least $1-1/e$.
  Conditioning on the existence of this hash function, the probability of a point projecting above $\Delta$ is at least $ 1-(1-1/\phi)^{2\phi}\geq 1-\frac{1}{e^2}$.
  Then it follows that $\Pr[|D\cap A(q,r,w)|< 1]< 1/e+1/e^2$.
  The points in $D$ will necessarily be added to $S_q$ before all other points in the buckets; 
  then, if we allow for $|S_q|=m+3L$, we get $$\Pr[|S_q\cap A(q,r,cw)|\geq 1]\geq1-(1/3+1/e+2/e^2)>0.02.$$\qed

  The data structure requires us to store the top $\BO{mL}$ points per projection vector,
  per bucket,
  for a total space cost of  $\BOx{m\ell L^2}$, in addition to storing the dataset, $\BOx{nd}$.
  The query time is $\BO{\ell L+m(d+\log\ell L)}$. 
  The first term is for initializing the priority queue, and the second for constructing $S_q$ and calculating distances.   
  Substituting in $L\in\mathcal{O}(n^{\rho})$ and $\ell,m\in\mathcal{O}(n^{1/c^2}\log^{(1-1/c^2)/2}n)$ we get query time:
\begin{equation}
  \BO{n^{\rho+1/c^2}\log^{\lambda}{n}+n^{1/c^2}
    \log^{\lambda}{n}\left(d+\log{(n^{\rho+1/c^2}
      \log^{\lambda}{n})}\right)} \, ,
\end{equation}
where $\lambda=(1-1/c^2)/(2)$. Depending on the parameters different terms might dominate the cost, but for large $d$ we can simplify
to the version stated in the theorem. The hash buckets take space:

\begin{equation}
\BO{n^{2(\rho+1/c^2)}\log^{1-1/c^2}{n}}.
\end{equation}

Depending on $c$, we might want to bound the space by $\BOx{n\ell L}$ instead, which  yields a bound of $\BOx{n^{1+\rho+1/c^2}\log^{(1-1/c^2)/2}{n}}$. \qed
\end{proof}


\section{Conclusions and future work}

We have proposed a data structure for AFN with theoretical and
experimental guarantees. 
We have introduced the approximate annulus query and given a theoretical sublinear time 
solution. Although we have proved that it is not possible to
use less than $\min\{n, 2^{\BOM{d}}\}-1$ total space for $c$-AFN when the $c$
approximation factor is less than $\sqrt{2}$, it is an open problem to close
the gap between this lower bound and the space requirements of our result. 
Another interesting problem is to apply our data structure to improve the
output sensitivity of near neighbor search based on
locality-sensitive hashing. By replacing each hash bucket with an AFN data structure with suitable
approximation factors, it is possible to control the number of times each
point in $S$ is reported.

Our data structure extends naturally to general metric spaces.  Instead of
computing projections with dot product, which requires a vector space, we
could choose some random pivots and order the points by distance to each
pivot.  The query operation would be essentially unchanged.  Analysis and
testing of this extension is a subject for future work.


\bibliographystyle{splncs03}
\bibliography{biblio}

\end{document}